\documentstyle[12pt]{article}

\begin{document}
\thispagestyle{empty}
\begin{flushright}
UT-818\\
June, 1998\\
\end{flushright}

\bigskip
\bigskip

\begin{center}
\noindent{\Large \bf Absorption of Fermions by D3-branes}\\
\bigskip
\bigskip
\noindent{\large Kazuo Hosomichi\footnote{
                 E-mail: hosomiti@hep-th.phys.s.u-tokyo.ac.jp}}\\
\bigskip
{\it Department of Physics, University of Tokyo, \\
\medskip
Tokyo 113, Japan}
\bigskip
\bigskip
\end{center}
\begin{abstract}
  The absorption cross section of dilatinos by D3-branes is calculated by
means of both classical type IIB supergravity and the effective gauge field
theory on their worldvolume. The two methods give the same results, supporting
the microscopic description of black holes in terms
of D-branes and giving another evidence of AdS/superconformal reciprocity.
\end{abstract}

\newpage

\section{Introduction}
\setcounter{equation}{0}
  String theory or M theory are believed to give the microscopic description
for black holes.
  One of the applications of this idea is to the Hawking process.
  Using the effective field theory on the intersection of D-branes or
other solitons one can evaluate microscopically the absorption or emission
of particles by black holes, which are semiclassically calculated by
solving the field equations in black hole background. 
  The success of this idea has lead to a new conjecture\cite{9711200}
that the large $N$ limit of certain superconformal field theories are
``dual'' to the string theory or M theory on the product space of 
anti de-Sitter space and sphere.

  In this paper we undertake the semiclassical and microscopic calculations
of the absorption cross section of dilatinos by parallel D3-branes, where the
absorption of scalars are calculated in some earlier 
works\cite{9702076,9703040,9708005,9803023}. 
  We first calculate it from the solution of the Dirac equation in 
ten-dimensional type IIB supergravity on D3-brane background.
  Then we calculate it microscopically from the two-point
function of certain operators in $D=4,N=4$ U(N) gauge theory on the
worldvolume of D3-branes.
  The results of two calculations agree, giving further support to the
idea of D-brane description of black holes and AdS/CFT correspondence.

\section{Semiclassical calculation}
\setcounter{equation}{0}
  In this section we calculate the absorption cross section by solving
the Dirac equation of dilatinos in type IIB supergravity.   
  The extremal D3-brane solution in this theory is\cite{Horowitz}
\begin{eqnarray}
  ds^2_{(10)}&=&H^{-\frac{1}{2}}(-dx_0^2+\cdots+dx_3^2)
               +H^{\frac{1}{2}}(dx_4^2+\cdots+dx_9^2) \\
  C^+_{0123}&=& H^{-1}
\end{eqnarray}
\begin{equation}
  H=1+\frac{4\pi gN}{r^4}= 1+\frac{R^4}{r^4}
\end{equation}  
and we quote the Dirac equation from \cite{Schwarz}:
\begin{equation}
  \Gamma^\mu D_\mu \lambda
 =\frac{i}{960}\Gamma^{\rho_1\cdots\rho_5}\lambda F_{\rho_1\cdots\rho_5}.
\end{equation}
  Inserting the D3-brane solution into this equation gives
\begin{equation}
\left[ H^{\frac{1}{2}}\Gamma^a\partial_a+\Gamma^i\partial_i
      +\frac{i}{4}\Gamma^{0123i}\partial_i(\ln H)\right]
(H^{\frac{1}{8}}\lambda)=0
\label{the-equation-to-solve}
\end{equation}
where $\Gamma^a,\Gamma^i$ are field-independent gamma-matrices and
$a=\left\{0,1,2,3\right\}, i=\left\{4,\ldots,9\right\}$ are indices
that run directions tangent or normal to the branes, respectively.
  We adopt $(-(+)^9)$ signature where $\Gamma^0$ is
anti-hermite and $\Gamma^1,\cdots,\Gamma^9$ are hermite.
  The matrix $\Gamma^{0123}$ has eigenvalues $\pm i$.

  Next we shall find the spherical wave solution to 
(\ref{the-equation-to-solve}).
  By the term `spherical wave' we mean the wave that is spherical with respect
to the six-dimensional spatial directions transverse to the branes.
  This can be done as in the case of lower space-time dimensions\cite{9711072}.
  Firstly we introduce the orbital angular momentum and spin operators
\[ L_{ij}=x_i\partial_j-x_j\partial_i\;,\;\;
   \Sigma_{ij}=\frac{1}{2}\Gamma_{ij}. \]
Then we put into (\ref{the-equation-to-solve}) the spherical wave
decomposition form for $\lambda$
\begin{equation}
 H^{\frac{1}{8}}\lambda=e^{-i\omega t}r^{-\frac{5}{2}}
 \left\{F(r)\cdot\Psi_{-l}+iG(r)\cdot
   (\frac{\Gamma^0\Gamma^i x_i}{r})\Psi_{-l}^{\pm}\right\}
\label{spherical}
\end{equation}
where $\Psi_{-l}^{\pm}$ is the eigenspinor of total angular momentum with
$(\Sigma_{ij}L_{ij})=-l,\Gamma^{0123}=\pm i$.
  We can set the spatial momenta along the branes equal to zero by Lorentz
transformations.
  We can easily obtain the radial wave equation
\begin{eqnarray}
  \omega H^{\frac{1}{2}}F+\left[\frac{d}{dr}+\frac{l+5/2}{r}
   \pm\frac{1}{4}(\ln H)'\right]G&=&0 \nonumber \\
 -\omega H^{\frac{1}{2}}G+\left[\frac{d}{dr}-\frac{l+5/2}{r}
   \mp\frac{1}{4}(\ln H)'\right]F&=&0 
  \label{F-and-G}
\end{eqnarray}
and the expression for conserved flux
\begin{equation}
  f\equiv i(F^* G-G^* F).
  \label{conserved-flux}
\end{equation}
  The sign $\pm$ in (\ref{F-and-G}) depends on whether we take
$\Psi_{-l}^{+}$ or $\Psi_{-l}^{-}$ in (\ref{spherical}) .

  We construct the approximate solution to (\ref{F-and-G})
with the minus sign by the following procedure.
  We can rewrite (\ref{F-and-G}) and (\ref{conserved-flux})
in terms of a new radial variable $x\equiv r/R$ and a new function $\phi$ as
\begin{equation}
  \left[ (x\frac{d}{dx})^2-(l+2)^2+\omega^2 R^2(x^2+x^{-2})\right]\phi=0
  \label{the-radial-wave-equation-in-terms-of-phi}
\end{equation}
\begin{equation}
  f=i(x\frac{d}{dx}\phi^\ast\cdot\phi-\phi^\ast\cdot x\frac{d}{dx}\phi).
\end{equation}
  In the asymptotic region or when $x\gg \omega R$ we can neglect the fourth
term in the L.H.S of (\ref{the-radial-wave-equation-in-terms-of-phi}) and the
approximate solution is expressed by the Bessel and Neumann functions as
\begin{equation}
  \phi=a J_{l+2}(\omega R x)+b N_{l+2}(\omega R x).
\end{equation}
  In the horizon region or when $\omega R x\ll 1$ the third term can be
neglected in turn and the solution becomes
\begin{equation}
  \phi=J_{l+2}(\omega R x^{-1})+iN_{l+2}(\omega Rx^{-1})
\end{equation}
up to multiplication by scalars.
  The solution is unique when the boundary condition is imposed so that $\phi$
has no outgoing flux on the horizon.
  By continuing these two solutions we can determine the coefficients $a,b$.
  The absorption cross section of plane wave can be calculated as follows:
\begin{eqnarray}
  \sigma_{\rm abs}&=&\frac{4\pi^2}{3\omega^5}\frac{(l+4)!}{l!}
  \cdot\frac{f_{\rm abs}}{f_{\rm in}} \nonumber \\
  &=&\frac{(l+4)!}{l![(l+1)!(l+2)!]^2}\frac{16\pi^4}{3\omega^5}
  \left(\frac{\omega R}{2}\right)^{4l+8}.
\label{absorption-cross-section}
\end{eqnarray}
  On the contrary, solving the equation (\ref{F-and-G}) with
the plus sign gives
\begin{equation}
  \sigma_{\rm abs}=\frac{(l+4)!}{l![(l+2)!(l+3)!]^2}\frac{16\pi^4}{3\omega^5}
  \left(\frac{\omega R}{2}\right)^{4l+12}
\end{equation}
which says that the absorption of these modes are suppressed by a
factor of $\omega^4$.
  In the s-wave case (\ref{absorption-cross-section}) becomes as
\begin{equation}
  \sigma_{\rm abs}=\frac{\pi^4\omega^3 R^8}{8}=\frac{G_{10}N^2\omega^3}{4}.
\label{semiclassical-result}
\end{equation}

\section{Microscopic Calculation}
\setcounter{equation}{0}

  In this section we shall calculate the absorption cross section
microscopically.
  To begin with, we quote the action describing the motion of a D3-brane
in generic type IIB supergravity background 
\cite{9610148,9611159,9611173}
\begin{eqnarray}
I &=& I_{\rm DBI}+I_{\rm WZ} \nonumber \\
  &=& -\int d^4 x
       \sqrt{-{\rm det}(g_{ij}+e^{-\frac{\phi}{2}}{\cal F}_{ij})}
      +\int e^{\cal F}\wedge {\cal C},  \label{DBI-action-for-D3brane}
\end{eqnarray}
where
\[ g_{ij}=E_i^a E_j^b \eta_{ab}\;,\;\;
   {\cal F}_{ij}=F_{ij}-B^{(\rm NS )}_{ij} \]
and ${\cal C}$ is the collection of RR forms.
  The pull-back of superforms is defined, for example, as
\begin{equation}
 B^{(\rm NS)}_{ij}\equiv E_j^B E_i^A B^{(\rm NS)}_{AB}.
\end{equation}
  Putting the on-shell superfield configuration into 
(\ref{DBI-action-for-D3brane}) and expanding it with respect to the fermionic
coordinate $\theta^\alpha, \bar{\theta}^{\bar{\alpha}}$, it becomes,
in the static gauge, as
\begin{eqnarray}
{\cal L}&=& -1+\frac{1}{2}\delta_{IJ}\partial X^I \partial X^J
            +\frac{i}{2}(\theta\sigma^i \partial_i\bar{\theta}
                        +\bar{\theta}\sigma^i\partial_i\theta)
            -\frac{1}{4}e^{-\phi}F_{ij}F^{ij}
            +\frac{1}{4}\chi F_{ij}\tilde{F}^{ij} \nonumber \\
        & & +({\rm nonrenormalizable\; terms}) \nonumber \\ 
        & & +({\rm terms\; containing\; the\; external\;\; fields}).
\label{Abelian-gauge-theory-action-for-a-single-D3brane}            
\end{eqnarray}
where $i,j=0,1,2,3$ and $I,J=4,\ldots,9$ are indices of tangential and normal
directions to the brane and
 $\sigma^m\;(m=0,1,\ldots,9)$ are $16\times 16$ matrices from which the 
ten-dimensional Dirac matrices are constructed.\footnote{
We follow largely the convention of \cite{Howe}. Definitions and formulas
which are relevant in this article are summarized in the appendix.}
  By the term `external fields' we mean the fields of type IIB 
supergravity in the bulk.
  
  The above action contains twice the fermionic degree of freedom as is
needed.
  We can reduce them by using $\kappa$-symmetry.
  The $\kappa$-transformation acts on the fermionic coordinates as
\[ \delta_{\kappa}Z^M E_M^\alpha=\kappa^\alpha\;,\;\;
   \delta_{\kappa}Z^M E_M^{\bar{\alpha}}={\bar{\kappa}}^{\bar{\alpha}} \]
and $\kappa$ satisfies, to the zeroth order in the world-volume fields,
the following conditions
\[ \Gamma\kappa=-i\kappa\;,\;\;
   \Gamma\bar{\kappa}=i\bar{\kappa}\]
\[ \Gamma= \frac{1}{4!\sqrt{-g}}\varepsilon^{ijkl}\hat{\sigma}_{ijkl} 
         = \hat{\sigma}_{0123} \]
  We shall hereafter call spinors with 
$\sigma_{0123}=\pm i\;{\rm or}\; \hat{\sigma}_{0123}=\pm i$
as the Left-(Right-)handed spinors in the four-dimensional sense.
  We take the following $\kappa$-gauge choice
\begin{equation}
  \theta^\alpha=\theta^\alpha_{\rm L} \;,\;\;
  \bar{\theta}^{\bar{\alpha}}=\bar{\theta}^{\bar{\alpha}}_{\rm R}.
\end{equation}

  The bulk dilatinos couple to the fields on the worldvolume as
\begin{eqnarray}
{\cal L}_{\rm int}&=&
   e^{-\frac{\phi_0}{2}}\theta_{\rm L}\sigma_{ij}F^{ij}\bar{\lambda}
  -e^{-\frac{\phi_0}{2}}\bar{\theta}_{\rm R}\sigma_{ij}F^{ij}\lambda
   \nonumber \\
  &=& {\cal O}_{[\bar{\lambda}]}^{\bar{\alpha}}\bar{\lambda}_{\bar{\alpha}}
     +{\cal O}_{[\lambda]}^\alpha \lambda_\alpha
\label{interaction-with-the-external-fields}
\end{eqnarray}
where $\phi_0$ is the VEV of dilaton.
  Here we have neglected the terms of higher order in the external fields
and the terms containing nonrenormalizable operators of the worldvolume
gauge theory.
  One can find these interactions by considering the following terms in 
(\ref{DBI-action-for-D3brane})
\[ \frac{1}{2}e^{-\phi}F^{ij}B^{\rm NS}_{ij}
  +\frac{1}{2}\tilde{F}^{ij} B^{\rm R}_{ij} \]
and extracting terms linear in $\theta,\bar{\theta}$.
  There is no other contributions to this order.
  One can easily find out from (\ref{interaction-with-the-external-fields})
that only the following components
\[ \lambda_{\rm R}\;,\;\;\bar{\lambda}_{\rm L} \]
couple to operators of dimension $7/2$ on the worldvolume.
  This is consistent with the semiclassical result that the absorption of
dilatinos are enhanced or suppressed according to the signs of their
interaction with five-form field strength. 

  As is discussed in \cite{9702015,9706100,9708005}, one can calculate 
the absorption cross section from the `discontinuity' of two point function
of the operator that couples to the relevant external field.
  Let us calculate it for the case of the absorption of dilatinos.
  Using the following propagators
\begin{eqnarray}
\left<X^I(x)X^J(0)\right>&=&-i\delta^{IJ}G_F(x) \nonumber \\
e^{-\phi_0}\left<A_i(x)A_j(0)\right>&=&i\eta_{ij}G_F(x) \nonumber \\
\left<\theta_{\rm L}^\alpha(x) \bar{\theta}_{\rm R}^{\bar{\alpha}}(0)\right>
&=&(P_L\hat{\sigma}^m P_R)^{\alpha\bar{\alpha}}\partial_m G_F(x) \nonumber \\
 G_F(x)&=& \frac{1}{4i\pi^2}\frac{1}{x^2}
\end{eqnarray}
we can easily calculate the two-point function
\begin{eqnarray}
\Pi_{[\lambda]}^{\alpha\bar{\alpha}}(x)&\equiv&
\left<{\cal O}_{[\lambda]}^\alpha(x)
        {\cal O}_{[\bar{\lambda}]}^{\bar{\alpha}}(0)\right> \nonumber \\
 &=& -4i(\hat{\sigma}^m)^{\alpha\bar{\alpha}}\partial_m\partial^2 G_F(x)^2.
\end{eqnarray}
  The Fourier transform of $\Pi_{[\lambda]}^{\alpha\bar{\alpha}}(x)$ is
defined as
\[ \Pi_{[\lambda]}^{\alpha\bar{\alpha}}(p)
   =\int d^4x\Pi_{[\lambda]}^{\alpha\bar{\alpha}}(x)e^{ipx} \]
and its `discontinuity across the real axis' is calculated as
\begin{eqnarray}
\Sigma_{[\lambda]}^{\alpha\bar{\alpha}}&\equiv&
\frac{1}{2i\omega}
\left[\Pi_{[\lambda]}^{\alpha\bar{\alpha}}(p)|_{p^0=\omega+i\epsilon,p^i=0}
     -\Pi_{[\lambda]}^{\alpha\bar{\alpha}}(p)|_{p^0=\omega-i\epsilon,p^i=0}
\right] \nonumber \\
 &=& \frac{N^2}{4\pi}(\hat{\sigma^0})^{\alpha\bar{\alpha}}\omega^2.
\end{eqnarray}
  In the above expression we have incorporated a factor $N^2$ that arises from
multiple D3-branes.
  In fact we do not know the precise form of the action for multiple
D3-branes, but the factor $N^2$ arises naturally when we assume that
the effective theory on the worldvolume of $N$ coincident D3-branes be
$U(N)$ gauge theory and adopt the symmetrized trace prescription of
\cite{9701125}.

  As for the bulk dilatino states, we write down the second quantized form
for $\lambda_\alpha,\bar{\lambda}_{\bar{\alpha}}$:
\begin{eqnarray}
 \lambda_\alpha(x)&=&\sqrt{\frac{16\pi G_{10}}{32}}
 \int\frac{d^6{\bf k}}{(2\pi)^6 2\omega}\sum_i
  [b({\bf k},i)u_\alpha({\bf k},i)e^{-ikx}
  +d^\dagger({\bf k},i)v_\alpha({\bf k},i)e^{ikx}] \nonumber \\
 \bar{\lambda}_{\bar{\alpha}}(x)&=&\sqrt{\frac{16\pi G_{10}}{32}}
 \int\frac{d^6{\bf k}}{(2\pi)^6 2\omega}\sum_i
  [b^\dagger({\bf k},i)u^\ast_{\bar{\alpha}}({\bf k},i)e^{ikx}
  +d({\bf k},i)v^\ast_{\bar{\alpha}}({\bf k},i)e^{-ikx}] \nonumber \\
& &
\end{eqnarray}
where $x$ and $k$ are seven-dimensional ones.
  The factor $(16\pi G_{10}/32)$ relates the dilatinos in \cite{Howe} to the
canonically-normalized ones.
  The canonical anticommutation relation reads
\[
\left\{ \lambda_\alpha(x),\bar{\lambda}_{\bar{\alpha}}(y)\right\}|_{x^0=y^0}
  =   (\frac{16\pi G_{10}}{32})
      \delta_{\alpha\bar{\alpha}}\delta^6({\bf x}-{\bf y}) \]
\begin{equation}
\left\{ b({\bf k},i), b^\dagger({\bf k'},j)\right\}=
\left\{ d({\bf k},i), d^\dagger({\bf k'},j)\right\}=
 (2\pi)^6 2\omega\delta^6({\bf k}-{\bf k'})\delta_{ij}
\end{equation}
and the spinor wave functions $u_\alpha({\bf k},i),v_\alpha({\bf k},i)$
satisfy
\[
\sum_\alpha
u^\ast_\alpha({\bf k},i)u_\alpha({\bf k},j)=
\sum_\alpha
  v^\ast_\alpha({\bf k},i)v_\alpha({\bf k},j)=2\omega\delta_{ij} \]
\[
\sum_\alpha
  u^\ast_\alpha({\bf k},i)v_\alpha({\bf -k},j)=
\sum_\alpha
  v^\ast_\alpha({\bf k},i)u_\alpha({\bf -k},j)=0 \] 
\begin{equation}
\sum_i u_\alpha({\bf k},i)u^\ast_{\bar{\alpha}}({\bf k},i)=
\sum_i v_\alpha({\bf k},i)v^\ast_{\bar{\alpha}}({\bf k},i)=
k_m(\sigma^m)_{\alpha\bar{\alpha}}
\end{equation}
  The absorption cross section of dilatinos $\sigma_{\rm abs}$ is
calculated as
\begin{eqnarray}
\sigma_{\rm abs}\delta_{ij}&=&
\left(\frac{16\pi G_{10}}{32}\right)\Sigma_{[\lambda]}^{\alpha\bar{\alpha}}
u^\ast_{\bar{\alpha}}({\bf k},i)u_\alpha({\bf k},j) \nonumber \\
&=& \delta_{ij}\frac{G_{10}N^2\omega^3}{4}.
\end{eqnarray}
  This agrees with the macroscopic result (\ref{semiclassical-result}).

\section{Discussions}
  We have calculated in this paper the absorption cross section of dilatinos
in two different ways.
  The two results agree, to the leading order in $\omega$, both presenting
the remarkable feature that the absorption is enhanced or suppressed
according to their signs of interaction with five-form field strength.

  Notice that in doing perturbative analysis of gauge theory we assumed
$gN\ll 1$ and dropped terms of higher order in $gN$ in calculating the 
two-point function.
  But this is exactly when $R$ or the size of the black 3-brane is much
smaller than the string scale and semiclassical approximation is not good.
  To see the correspondence of gauge theory and supergravity one should
properly treat the non-abelian nature of the worldvolume theory and
analyze the theory with large $gN$.
  Our result suggests that the value of the two point function 
$\left<{\cal O}_{[\lambda]}^\alpha(x)
        {\cal O}_{[\bar{\lambda}]}^{\bar{\alpha}}(0)\right>$
remain the same at large $gN$.
  This does not seem obvious to us.

  Checking the coincidence for the gravitinos and particles with other spin
would be interesting.
  It seems that microscopic evaluation of the absorption of gravitinos
is much more involved than in the case of dilatinos.
  One reason for this is the $\kappa$-symmetry.
  Expanding the D3-brane action (\ref{DBI-action-for-D3brane})with respect to
$\theta$, one can find that the interaction terms by which the gravitinos
couple to the worldvolume operators of the lowest dimension are
\[ (i\theta\sigma_i\bar{\psi}^i+i\bar{\theta}\sigma_i\psi^i)
  +\frac{1}{6}\epsilon^{ijkl}(\bar{\theta}\sigma_{ijk}\psi_l
   -\theta\sigma_{ijk}\bar{\psi}_l).\]
  These terms come from $I_{\rm DBI}$ and $\int C_{0123}$, respectively.
  With our gauge-fixing these two terms cancel, but for other gauge choices
such as proposed in \cite{9705040} these terms don't vanish in general.
  It is somewhat strange, since the coupling of bulk fields and worldvolume
fields seems $\kappa$-gauge-dependent, although the above interaction may yield
no discontinuity to the two-point function and is invisible by the analysis of
absorption cross section.
  Anyway we believe that a careful analysis will lead to a $\kappa$-invariant
expression for absorption cross section of gravitinos as in the case of
dilatinos.

  A recent work\cite{9805140} revealed that the field equation of a minimal
scalar in D3-brane background can be exactly solved in terms of Mathieu
functions.
  It seems that by applying this observation to the fermionic case one can
get further implications on the world-volume theory of D3-branes.

  Some other recent works\cite{9804149,9805082} claim that one can obtain
all the Kaluza-Klein modes of AdS supergravity by expanding D3-brane action
around AdS background.
  It would be interesting in this regard to see how higher-dimensional
fermionic operators work out.

\medskip

  The author thanks T. Eguchi for introducing to this subject,
Y. Sugawara, T. Kawano and K. Okuyama for encouragements and discussions.
  The author is also grateful for S. Das for communication and discussion
on recent issues.
  The author is supported in part by JSPS Research Fellowships for 
Young Scientists.

\appendix

\section{Spinors in ten dimensions}
\setcounter{equation}{0}
  We follow and use the convention of \cite{Howe}.
  We use the representation in which the Gamma matrix can be expressed as
\begin{eqnarray}
\Gamma^a&=&\left(\begin{array}{cc}0 & (\sigma^a)_{\alpha\beta} \\
                          (\hat{\sigma}^a)^{\alpha\beta} & 0
           \end{array}\right) \nonumber \\
\Gamma^{11}&=& \left(\begin{array}{cc} 1&0\\0&-1 \end{array}\right)
\end{eqnarray}
where the $16\times 16$ matrices $\sigma^a,\hat{\sigma}^a$ are real symmetric
and satisfy the following identity:
\[ (\sigma^a)=(1,\sigma^i)\;,\;\;(\hat{\sigma}^a)=(1,-\sigma^i)\;,\;\;\;\;
   i=1,\ldots,9\]
\[ \left\{ \sigma^i,\sigma^j \right\}=2\delta^{ij}.\]
Antisymmetric products of $\sigma$-matrices are defined as
\[\sigma^{ab\cdots c}=\sigma^{[a}\hat{\sigma}^b\cdots^{c]}\;\;,\;\;
  \hat{\sigma}^{ab\cdots c}=\hat{\sigma}^{[a}\sigma^b\cdots^{c]}. \]

  Ten-dimensional Dirac spinor has 32 components and according to the 
eigenvalue of $\Gamma^{11}$ it is decomposed into left-handed and right-handed
spinors:
\begin{eqnarray}
\psi&=&\psi_L+\psi_R \nonumber \\
\psi_L&=&\frac{1}{2}(1+\Gamma^{11})\psi=
\left(\begin{array}{c}\varphi_\alpha \\ 0\end{array}\right) \nonumber \\
\psi_R&=&\frac{1}{2}(1-\Gamma^{11})\psi=
\left(\begin{array}{c} 0\\ \chi^\alpha \end{array}\right)
\end{eqnarray}


\newpage

\begin{thebibliography}{99}

\bibitem{9711200}
J. Maldacena,
``The Large N Limit of Superconformal Field Theories and Supergravity'',
hep-th/9711200.

\bibitem{9702076}
I. Klebanov,
Nucl. Phys. B496 (1997) 231, hep-th/9702076.

\bibitem{9703040}
S. Gubser, I. Klebanov and A. Tseytlin,
Nucl. Phys. B499 (1997) 217, hep-th/9703040.

\bibitem{9708005}
S. Gubser and I. Klebanov,
Phys. Lett. B413 (1997) 41, hep-th/9708005.

\bibitem{9803023}
S. Gubser, A. Hashimoto, I. Klebanov and M. Krasnitz,
hep-th/9803023.

\bibitem{Horowitz}
G. Horowitz and A. Strominger,
Nucl. Phys. B360 (1991) 197.

\bibitem{Schwarz}
J. Schwarz,
Nucl. Phys. B226 (1983) 269.

\bibitem{9711072}
K. Hosomichi,
hep-th/9711072.

\bibitem{9610148}
M. Cederwall, A. von Gussich, B. Nilsson and A. Westerberg,
Nucl.Phys. B490 (1997) 163, hep-th/9610148.

\bibitem{9611159}
M. Cederwall, A. von Gussich, B. Nilsson, P. Sundell and A. Westerberg,
Nucl.Phys. B490 (1997) 179, hep-th/9611159.

\bibitem{9611173}
E. Bergshoeff and P. K. Townsend,
Nucl.Phys. B490 (1997) 145, hep-th/9611173.

\bibitem{Howe}
P. Howe and P. West,
Nucl. Phys. B238 (1984) 181.

\bibitem{9702015}
J. Maldacena and A. Strominger,
Phys. Rev. D56 (1997) 4975, hep-th/9702015.

\bibitem{9706100}
S. Gubser,
Phys. Rev. D56 (1997) 7854, hep-th/9706100.

\bibitem{9701125}
A. Tseytlin,
Nucl. Phys. B501 (1997) 41, hep-th/9701125.

\bibitem{9705040}
E. Bergshoeff, R. Kallosh, T. Ortin and G. Papadopoulos,
Nucl. Phys. B502 (1997) 149, hep-th/9705040.

\bibitem{9805140}
S. Gubser and A. Hashimoto, hep-th/9805140.

\bibitem{9804149}
S. Das and S. Trivedi, hep-th/9804149.

\bibitem{9805082}
S. Ferrara, M. Lledo and A. Zaffaroni, hep-th/9805082.

\end{thebibliography}
\end{document}